\documentclass[11pt]{article}

\PassOptionsToPackage{hyphens}{url}
\usepackage[final]{acl}

\usepackage{times}
\usepackage{latexsym}
\usepackage[T1]{fontenc}
\usepackage[utf8]{inputenc}
\usepackage{microtype}
\usepackage{inconsolata}
\usepackage{graphicx}

\usepackage{booktabs}
\usepackage{xspace}
\usepackage{amsmath,amssymb}
\usepackage{array}
\usepackage{algorithm}
\usepackage{algorithmic}
\usepackage{fvextra}
\usepackage{float}
\usepackage{placeins}
\newcommand{\tool}{EffiSkill\xspace}
\hypersetup{
    pdfauthor={Zimu Wang; Yuling Shi; Mengfan Li; Zijun Liu; Jie M. Zhang; Chengcheng Wan; Xiaodong Gu},
    pdftitle={Automated Code Efficiency Optimization via Agent Skill Mining},
    pdfsubject={Code efficiency optimization via reusable agent skills},
    pdfkeywords={Code optimization, large language models, agent skill mining}
}

\begin{document}

\title{Automated Code Efficiency Optimization via Agent Skill Mining}

\author{
    Zimu Wang\textsuperscript{1,2}\thanks{Equal contribution.}\quad
    Yuling Shi\textsuperscript{1}\footnotemark[1]\quad
    Mengfan Li\textsuperscript{1}\quad
    Zijun Liu\textsuperscript{3}\\
    \textbf{Jie M. Zhang}\textsuperscript{4}\quad
    \textbf{Chengcheng Wan}\textsuperscript{5,6}\quad
    \textbf{Xiaodong Gu}\textsuperscript{1}\thanks{Corresponding author.}\\[0.5ex]
    {\normalfont\large \textsuperscript{1}Shanghai Jiao Tong University\quad
    \textsuperscript{2}University of California, Berkeley}\\
    {\normalfont\large \textsuperscript{3}Columbia University\quad
    \textsuperscript{4}King's College London}\\
    {\normalfont\large \textsuperscript{5}East China Normal University\quad
    \textsuperscript{6}Shanghai Innovation Institute}\\[0.5ex]
    {\normalfont\large\texttt{\{yuling.shi,lmf2951510526,xiaodong.gu\}@sjtu.edu.cn}}\\
    {\normalfont\large\texttt{zm.wang@berkeley.edu, zl3031@columbia.edu}}\\
    {\normalfont\large\texttt{jie.zhang@kcl.ac.uk, ccwan@sei.ecnu.edu.cn}}
}

\maketitle

\begin{abstract}
Code efficiency is a fundamental aspect of software quality, yet how to use large language models (LLMs) to optimize correct programs remains challenging. Prior approaches rely on one-shot rewriting, retrieved exemplars or instructions, fine-tuning, or candidate search.
In this paper, we present \tool, a two-stage framework that models recurring slow-to-fast transformations as \emph{Operator Skills}: persistent cards that record a transformation's intent, applicability, procedure, expected effect, and risks. Stage~I distills frozen, language-specific skill libraries from slow/fast program pairs. Stage~II diagnoses a new program, retrieves relevant skills, composes alternative plans, and generates up to eight candidates. These steps finish without candidate-execution feedback; public tests then filter and rank candidates, and held-out private tests determine reported correctness and speedup.
Across 623 EffiBench-X tasks per language, two backbones, and separate Python and C++ evaluations, \tool achieves the highest observed OPT@1 and OPT@8 among the compared methods in all four settings. Its OPT@8 margins over the strongest baselines range from 4.98 to 12.52 percentage points. These results support broader successful optimization coverage for the complete skill-guided pipeline under a shared output budget.
\end{abstract}

\begin{figure}[t]
\centering
\includegraphics[width=\columnwidth]{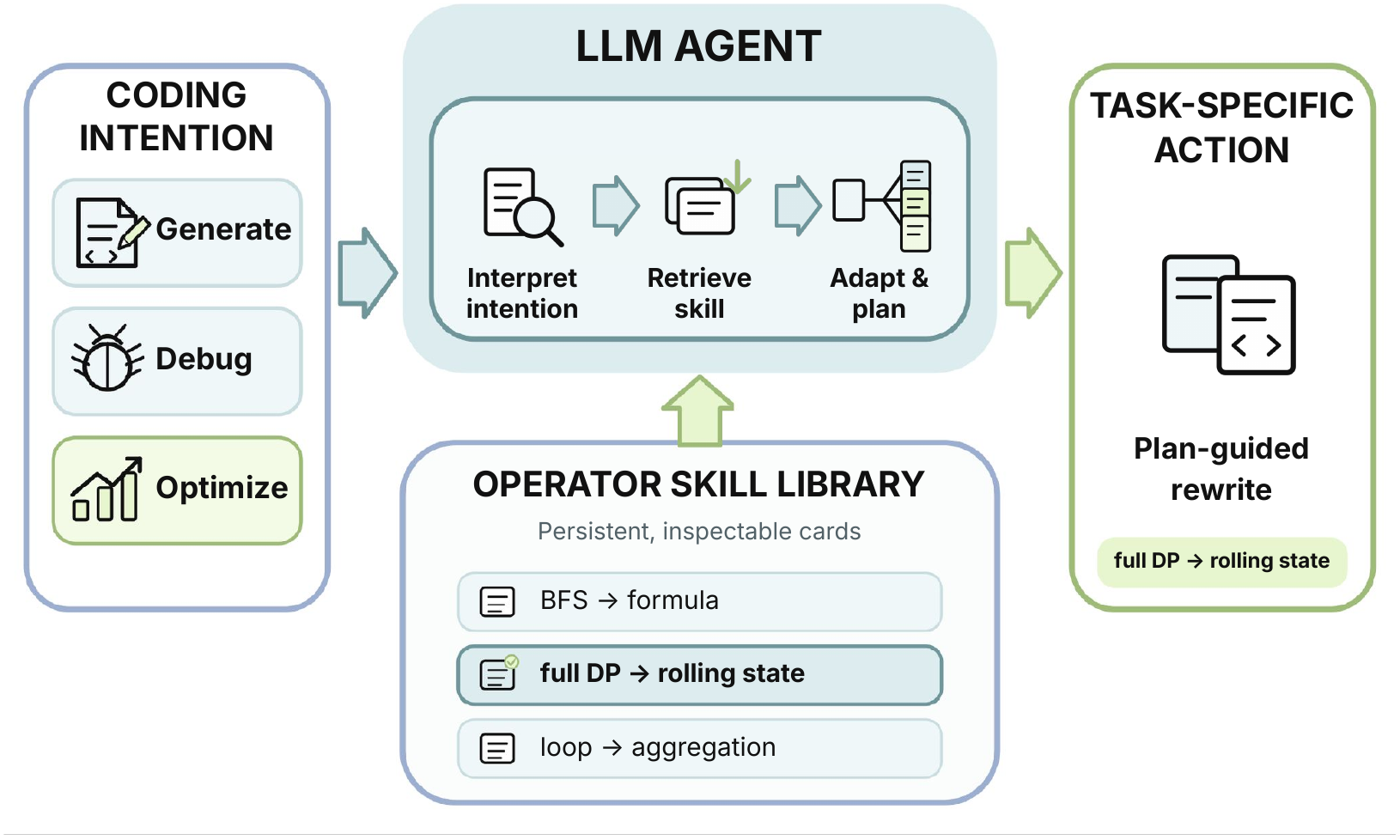}
\caption{Overview of an agent skill as reusable procedural knowledge.}
\label{fig:agent-skill-concept}
\end{figure}

\section{Introduction}\label{sec:intro}

The computational efficiency of software is a fundamental aspect of software quality~\cite{DBLP:conf/sigsoft/GargMCSW22}. Inefficient implementations increase latency and waste computational resources~\cite{DBLP:conf/msr/NistorJT13,DBLP:conf/oopsla/JovicAH11}, yet improving them is difficult: an optimization must preserve externally visible behavior while changing the computation responsible for its cost. Automatically improving a correct but inefficient program has therefore become an important problem in software engineering.

Early code-optimization systems mainly relied on manually designed rules for particular inefficiency patterns~\cite{DBLP:conf/msr/NistorJT13,nistor2015caramel}. More recently, LLM-based approaches have generated optimized code through direct prompting, retrieval augmentation, fine-tuning, or search-based refinement~\cite{shypula2023learning,SBLLM,DBLP:journals/jmlr/IzacardLLHPSDJRG23,huang2024efficoder}. Although these methods show encouraging progress, they leave an important gap between instance-level optimization and reusable optimization knowledge.

A central limitation is that many existing methods primarily operate at the level of individual examples. A prompting method rewrites one program; retrieval supplies another program with incidental details; fine-tuning distributes transformations across model weights; and a search trajectory records decisions for one task~\cite{SBLLM}. In these instance-centric approaches, the optimization mechanism is not available as a persistent, separately retrievable unit that states when it applies, how to perform it, and what risks must be checked.

\begin{figure}[t]
\centering
\fbox{%
\begin{minipage}{0.92\columnwidth}
\footnotesize
\textbf{O016: Algebraic Counting}\par
\emph{Replace cubic triple enumeration with lower-dimensional counting.}\par\smallskip
\begin{tabular}{@{}>{\raggedright\arraybackslash}p{0.43\linewidth}>{\raggedright\arraybackslash}p{0.43\linewidth}@{}}
\textbf{Applicability} & \textbf{Procedure} \\
Linear equality; counting rather than listing; the third variable is constrained by bounds. &
Fix two variables, derive or count the valid range of the third, and tighten feasible bounds. \\[2pt]
\textbf{Expected effect} & \textbf{Risks} \\
Typically \(O(n^3)\!\rightarrow\!O(n^2)\) with \(O(1)\) extra space. &
Off-by-one bounds, lost ordering constraints, non-linear predicates, or explicit-enumeration requirements.
\end{tabular}
\end{minipage}}
\caption{Compact form of Python Operator Skill O016. Appendix~\ref{app:inventory} provides the full card and library inventory.}
\label{fig:operator-skill-card}
\end{figure}

This instance-centric design is restrictive because performance-improving edits often exhibit recurring transformation patterns. Different programs may benefit from replacing repeated enumeration with aggregation, reformulating a computation algebraically, compressing dynamic-programming state, or reducing constant-factor overhead through better data structures. Treating every task independently misses this recurring structure.

In this paper, we propose \tool, a framework that addresses this gap by explicitly modeling code optimization through reusable \emph{Operator Skills}. Figure~\ref{fig:agent-skill-concept} presents the general notion of a skill connecting intent to action through procedural knowledge. In \tool, each skill is a structured unit distilled from recurring slow-to-fast transformations. The reusable unit is therefore a transformation mechanism rather than a raw program pair or implicit model behavior.

\tool follows a two-stage design. \textbf{Stage I (Skill Mining)} operates offline on paired slow and fast programs to distill frozen, language-specific skill libraries. \textbf{Stage II (Skill-Guided Optimization)} uses fixed orchestration to diagnose a new program, retrieve applicable Operator Skills, compose alternative task-specific plans, and generate candidates. Diagnosis through pre-execution processing finishes without candidate execution feedback. Public tests are then used to filter and rank the completed candidate pool, while held-out private tests determine the reported results.

We evaluate \tool on 623 EffiBench-X~\cite{qing2025effibench-x} tasks per language with GPT-5-mini and Qwen3-Coder in separate Python and C++ settings. \tool attains the highest observed optimization coverage among the compared methods in all four settings. At OPT@8, its margins over the strongest baselines range from 4.98 to 12.52 percentage points. We further analyze matched ablations, speedup magnitude, terminal outcomes, the learned library, and one complete end-to-end example.\footnote{Code and archived reproduction assets: \url{https://github.com/zimuwang-real/EffiSkill}.}

We summarize our contributions as follows:
\begin{enumerate}
    \item We introduce Operator Skills, explicit and inspectable representations distilled from recurring slow-to-fast program transformations.
    \item We propose \tool, a two-stage framework for offline skill mining and fixed skill-guided optimization.
    \item We evaluate \tool with two backbones in separate Python and C++ settings, covering matched ablations, coverage, speedup, learned skills, and one end-to-end example.
\end{enumerate}

\begin{figure*}[t]
\centering
\includegraphics[width=.72\textwidth]{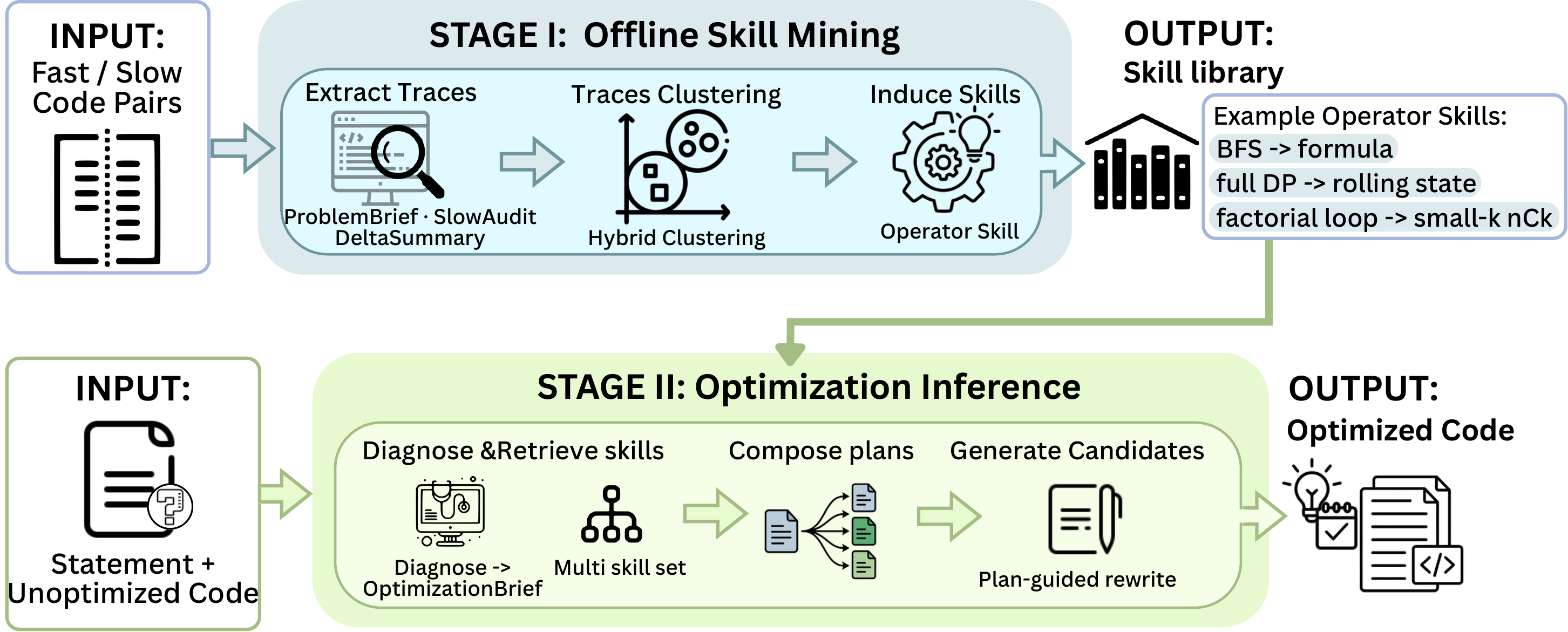}
\caption{\tool workflow. Stage I distills Operator Skills from slow/fast pairs; Stage II applies a frozen within-language registry through fixed diagnosis, retrieval, planning, generation, and pre-execution processing.}
\label{fig:overview}
\end{figure*}

\section{Background}\label{sec:background}

\subsection{Problem Definition}\label{subsec:problem}

Following prior work~\cite{shypula2023learning,SBLLM}, we study code-efficiency optimization for a functionally correct but inefficient program. Let \(S\) denote the input program and \(C\) the available task context, such as a problem statement or specification. The objective is to generate candidates \(\mathcal O=\{O_1,\ldots,O_m\}\) that preserve the intended behavior of \(S\) while improving efficiency. Let \(\mathcal T\) be an evaluation test suite and \(u(P)\) the benchmark's aggregate runtime for program \(P\). A successful candidate \(O\) must pass \(\mathcal T\) and satisfy \(u(O)<u(S)\); our reported metric applies the stricter threshold \(u(S)/u(O)\geq1.10\).

The optimization and evaluation phases are separate. Diagnosis, retrieval, planning, generation, and pre-execution processing finish without candidate execution feedback. Public tests then filter incorrect candidates and rank survivors by runtime; held-out private tests determine reported correctness and speedup. This boundary studies knowledge-guided generation without execution inside its loop while preserving the benchmark protocol used to select and evaluate outputs.

\subsection{Operator Skills}\label{subsec:skills}

Recent agent frameworks conceptualize skills as modular units that encapsulate instructions and resources for recurring workflows~\cite{schick2023toolformer,yao2022react,li2026skillsbench}. We adopt a definition tailored to code-efficiency optimization: an \emph{Operator Skill} is a persistent, separately retrievable card that records a transformation's intent, applicability, procedure, expected complexity or runtime effect, and risks. Compared with a raw example or an implicit parameter update, the card exposes both the reusable mechanism and the checks required to adapt it.

Figure~\ref{fig:operator-skill-card} makes this abstraction concrete. O016 replaces cubic triple enumeration with lower-dimensional counting when one variable is constrained by a linear relation and bounds. Its fields connect a recognizable trigger to transformation steps, an expected \(O(n^3)\!\rightarrow\!O(n^2)\) effect, and correctness risks. Operator Skills are the sole mined knowledge artifact carried from Stage~I into Stage~II; diagnosis, retrieval, planning, generation, and pre-execution processing are fixed orchestration around the frozen cards.

\section{Method}\label{sec:approach}

The key idea of \tool is to treat code optimization not as one-shot rewriting, but as a process of composing reusable transformation mechanisms learned from prior examples. Its two stages decouple learning optimization knowledge from applying that knowledge to a new program. Stage~I distills recurring slow-to-fast transformations into a frozen registry; Stage~II retrieves and adapts those mechanisms through diagnosis, alternative plans, and bounded candidate generation. Figure~\ref{fig:overview} summarizes the workflow.

The registry preserves transformations across tasks, while diagnosis and plans supply task-specific context and explore several coherent applications within the bounded candidate pool. The reusable mechanism remains distinct from the program-specific decisions needed to apply it.

\subsection{Stage I: Operator-Skill Distillation}\label{subsec:stage1}

Python and C++ use separate within-language mining lineages. The representation progression is slow/fast pairs \(\rightarrow\) structured traces \(\rightarrow\) signatures \(\rightarrow\) hybrid clustering \(\rightarrow\) provisional cards \(\rightarrow\) consolidation \(\rightarrow\) a frozen language-specific registry.

\paragraph{Pair normalization and trace extraction.}
Each language contributes 900 unique slow/fast comparisons, canonicalized and deduplicated before trace extraction. An LLM then produces a structured trace: \texttt{ProblemBrief} records the task and inferred constraints, \texttt{SlowAudit} identifies dominant operations and bottlenecks, \texttt{FastAudit} characterizes the faster mechanism and complexity, and \texttt{DeltaSummary} records the transformation, steps, triggers, and pitfalls.

\paragraph{Signature abstraction.}
Deterministic extraction forms a textual signature from the transformation type, before/after complexity, trigger signals, transformation steps, bottleneck kinds, and task tags. This removes incidental identifiers, statement wording, and output formatting while retaining the computational properties used to compare mechanisms.

\paragraph{Hybrid clustering.}
The Stage-I pipeline combines normalized TF--IDF unigram/bigram features with normalized \texttt{all-MiniLM-L6-v2} embeddings~\cite{all_minilm_l6_v2_2020}. Its KMeans pathway searches candidate cluster counts around \(\sqrt n\), evaluates them with cosine silhouette, and uses seed 42. Appendix~\ref{app:stage1} documents feature construction, cluster-count selection, normalization, and the complete construction procedure.

\paragraph{Card construction and consolidation.}
An LLM summarizes each selected cluster into representative transformations, triggers, complexity changes, procedures, and pitfalls. Programmatic consolidation uses scope compatibility, trigger and tag overlap, and textual similarity before rendering cards and serializing the frozen registry. We describe the process as \emph{automated, LLM-assisted procedural-knowledge distillation}: LLMs produce traces and cluster summaries, while signatures, feature construction, consolidation bookkeeping, card rendering, and serialization are programmatic. Appendix~\ref{app:stage1} specifies the complete Stage-I procedure, prompt and schema contract, deterministic post-processing, clustering configuration, and consolidation rules.

\subsection{Stage II: Skill-Guided Optimization}\label{subsec:stage2}

Stage II applies the mined skill library to unseen optimization tasks. Rather than attempting to identify a single correct rewrite upfront, \tool explores multiple plausible optimization routes through a structured pipeline consisting of diagnosis, retrieval, planning, and candidate generation.

\paragraph{Diagnosis and retrieval.}
Fixed orchestration first emits a structured diagnosis containing inferred constraints, bottlenecks, baseline complexity, dominant operations, and risks. Retrieval uses these signals to select applicable cards from the frozen, configuration-specific registry and organize them into one or more mechanism sets. This retrieval process is deliberately designed to return multiple alternatives, as inefficient programs often admit several valid optimization directions, and committing to a single skill choice too early can unduly narrow the search space.

\paragraph{Mechanism sets and plan composition.}
The planner interprets each mechanism set against \(S\), \(C\), and the diagnosis, then composes task-specific alternatives that record a proposed idea, target complexity, transformation steps, and risks. Alternative plans represent different applications, combinations, orders, or aggressiveness levels of the retrieved mechanisms.

\paragraph{Slot allocation, generation, and pre-execution processing.}
The controller allocates at most eight output slots across the available plans and requests implementations under the original interface contract. Before execution, language-specific processing records static, interface, syntax, or compiler diagnostics. Appendix~\ref{app:stage2} gives the schemas, prompt contracts, parsing and normalization, bounded allocation, static checks, and precise pre-execution boundary.

\paragraph{Public selection and private reporting.}
Public tests execute the completed candidate pool, filter incorrect candidates, and rank survivors by aggregate public runtime. Held-out private tests then determine reported correctness and speedup. Appendices~\ref{app:stage1} and~\ref{app:stage2} provide complete Stage-I and Stage-II pseudocode.

\section{Experimental Setup}\label{sec:evaluation}

\subsection{Benchmarks and Models}\label{subsec:protocol}

EffiBench-X~\cite{qing2025effibench-x} supplies 623 competitive-programming optimization tasks in each evaluated language. Each task's 100 tests are randomly partitioned into 20 public and 80 held-out private cases using a fixed per-task seed. We evaluate GPT-5-mini, snapshot \texttt{2025-08-07}~\cite{gpt5mini}, and Qwen3-Coder-30B-A3B-Instruct~\cite{qwen30b} in separate Python and C++ settings with language-specific frozen registries.

\subsection{Baselines}

We compare \tool with baselines from four categories.

\paragraph{Prompting.}
\textbf{Instruction} directly prompts the LLM to optimize code without intermediate reasoning or retrieval. \textbf{CoT}~\cite{wei2022chain} adds explicit bottleneck reasoning before generating optimized code.

\paragraph{Retrieval-augmented.}
\textbf{RAG} retrieves related slow/fast examples as few-shot optimization demonstrations~\cite{lewis2020retrieval,shypula2023learning}. \textbf{FasterPy} augments generation with optimization summaries retrieved from historical code transformations~\cite{wu2025fasterpy}.

\paragraph{Evolutionary search.}
\textbf{SBLLM} uses its archived initial-generation outputs under the shared eight-candidate budget~\cite{SBLLM}.

\paragraph{Fine-tuning.}
\textbf{EffiCoder}~\cite{huang2024efficoder} fine-tunes on execution-validated slow/fast pairs to directly generate optimized code; we evaluate its Qwen-based optimization checkpoint.

\subsection{Evaluation Protocol and Metrics}

Every method receives the same canonical program for a task and is evaluated under a shared eight-candidate output bound. Generation and pre-execution processing finish before evaluation. Public tests filter and rank generated candidates; private tests determine reported correctness and speedup.

\paragraph{Metric and statistical analysis.}
Let \(\mathcal Q\) be the 623 tasks in one model--language setting. For task \(q\), let \(S_q\) be the input program and let \(c_{q,1},\ldots,c_{q,m_q}\) be the survivors from one public-ranked \(K=8\) pool. Let \(u_q(P)\) be private \texttt{runtime\_sum} for program \(P\), and let \(\mathrm{pass}_{\rm priv}(P)\) indicate private-test correctness. Task-level success is
\begin{equation}
\begin{aligned}
z_q(k)=\mathbf{1}\!\Big[&
\exists j\leq\min(k,m_q):\mathrm{pass}_{\rm priv}(c_{q,j})\\
&{}\land u_q(S_q)/u_q(c_{q,j})\geq1.10\Big].
\end{aligned}
\label{eq:task-success}
\end{equation}
Aggregate optimization coverage is
\begin{equation}
\mathrm{OPT@}k=\frac{100}{|\mathcal Q|}
\sum_{q\in\mathcal Q}z_q(k).
\label{eq:opt}
\end{equation}
Thus OPT@1 and OPT@8 are prefixes of the same generated pool, and empty pools contribute zero. We compare each endpoint with its strongest observed baseline using a one-sided paired task bootstrap with 10,000 replicates and percentile 95\% confidence intervals on the paired mean difference. Holm adjustment across the eight endpoint p-values controls family-wise inference; Table~\ref{tab:main} marks adjusted \(p<.05\). Appendix~\ref{app:evaluation} reports the complete statistical table and protocol details.

To keep the paper self-contained, the appendices provide the complete method schemas, evaluation details, and reproducibility information needed to interpret the study.

\setcounter{topnumber}{2}
\section{Experimental Results}\label{sec:results}

\subsection{Overall Effectiveness}\label{subsec:main-results}

\begin{figure}[t]
\centering
\includegraphics[width=\columnwidth]{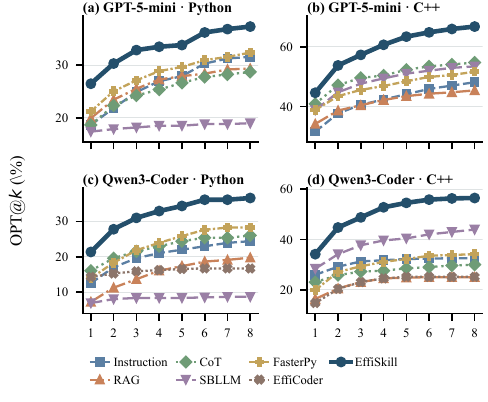}
\caption{OPT@\(k\) across successive prefixes of one public-ranked \(K=8\) candidate pool. Increasing \(k\) reveals more outputs from the same pool.}
\label{fig:topk_growth}
\end{figure}

\begin{table}[t]
\centering
\small
\setlength{\tabcolsep}{3pt}
\begin{tabular}{@{}lrrrr@{}}
\toprule
\multicolumn{5}{c}{\textbf{GPT-5-mini}} \\
& \multicolumn{2}{c}{\textbf{Python}} & \multicolumn{2}{c}{\textbf{C++}} \\
\cmidrule(lr){2-3}\cmidrule(l){4-5}
Method & OPT@1 & OPT@8 & OPT@1 & OPT@8 \\
\midrule
Instruction & 18.62 & 31.62 & 31.62 & 48.15 \\
RAG & 19.90 & 29.37 & 34.19 & 45.43 \\
CoT & 18.62 & 28.73 & 40.93 & 54.74 \\
SBLLM & 17.34 & 18.94 & 38.84 & 53.45 \\
FasterPy & 21.19 & 32.42 & 38.84 & 51.69 \\
\textbf{EffiSkill} & \textbf{26.48$^{*}$} & \textbf{37.40$^{*}$} & \textbf{44.62} & \textbf{66.77$^{*}$} \\
\midrule
\multicolumn{5}{c}{\textbf{Qwen3-Coder}} \\
& \multicolumn{2}{c}{\textbf{Python}} & \multicolumn{2}{c}{\textbf{C++}} \\
\cmidrule(lr){2-3}\cmidrule(l){4-5}
Method & OPT@1 & OPT@8 & OPT@1 & OPT@8 \\
\midrule
Instruction & 12.52 & 24.40 & 26.00 & 32.58 \\
RAG & 7.22 & 19.74 & 16.37 & 25.04 \\
CoT & 16.05 & 26.00 & 23.11 & 30.02 \\
SBLLM & 6.90 & 8.67 & 28.41 & 43.98 \\
FasterPy & 13.96 & 28.25 & 19.90 & 34.19 \\
EffiCoder & 14.29 & 16.69 & 14.77 & 25.20 \\
\textbf{EffiSkill} & \textbf{21.35$^{*}$} & \textbf{36.60$^{*}$} & \textbf{34.19$^{*}$} & \textbf{56.50$^{*}$} \\
\bottomrule
\end{tabular}
\caption{Optimization coverage (\%). \(^{*}\) denotes Holm-adjusted \(p<.05\) against the strongest observed baseline for that endpoint.}
\label{tab:main}
\end{table}

\tool has the highest observed OPT@1 and OPT@8 among the compared methods in all four model--language settings (Table~\ref{tab:main}). Its OPT@8 margins over the strongest observed baselines are 4.98 and 12.04 percentage points for GPT-5-mini on Python and C++, and 8.35 and 12.52 points for Qwen3-Coder. Python and C++ provide separate within-language evaluations with language-specific skill libraries.

This result is particularly important in our setting, where the central challenge is not merely to generate candidate rewrites, but to convert a limited candidate budget into successful efficiency improvements while preserving correctness.

Paired-bootstrap 95\% confidence intervals for the four OPT@8 margins are [0.64, 9.31], [7.54, 16.53], [3.69, 13.00], and [7.54, 17.50] percentage points for GPT/Python, GPT/C++, Qwen/Python, and Qwen/C++, respectively. Seven of the eight endpoint comparisons remain significant at \(\alpha=.05\) after Holm adjustment; GPT/C++ OPT@1 is the sole nonsignificant endpoint.

Figure~\ref{fig:topk_growth} shows that the observed coverage advantage persists across candidate-pool prefixes. Each \(k\) exposes another prefix of the same pool and measures the additional coverage supplied by later public-ranked candidates.

Moving from the first public-ranked candidate to the full eight-candidate pool increases \tool's coverage by 10.91, 22.15, 15.25, and 22.31 percentage points for GPT/Python, GPT/C++, Qwen/Python, and Qwen/C++, respectively, a range of 10.91--22.31 points. The largest increments occur in C++, where later candidates add more than 22 points for both backbones. Thus the bounded pool contributes materially beyond its leading candidate, consistent with alternative task-specific plans exploring several coherent optimization routes. The prefix curve reflects candidate-set diversity, not compute-matched output positions.

\subsection{Ablation Study}\label{subsec:analysis}

The matched Python suite evaluates three variants. \textbf{No retrieval} removes skill retrieval and generates candidates directly from the diagnosis. \textbf{Random cards} replaces retrieved cards with cards sampled from the library. \textbf{Single plan} retains diagnosis and retrieval but restricts generation to a single plan.

Relative to Full, removing retrieval lowers OPT@8 by 9.95 and 27.45 points for GPT-5-mini and Qwen3-Coder, respectively. Random cards reduce OPT@8 by 9.58 and 22.79 points, and the single-plan variant reduces it by 11.56 and 22.31 points. The common direction associates bottleneck-relevant retrieval and alternative task-specific plans with broader coverage in the tested system; the larger Qwen3-Coder drops show that sensitivity to this procedural context varies by backbone. These differences caution against treating either association as backbone-invariant.

Read together, the prefix growth and single-plan ablation are consistent with a benefit from maintaining several task-specific routes: later candidates add coverage, while reduced plan diversity is associated with lower coverage in the Python ablation. All ablated variants also fall below Instruction at OPT@8; because they differ from Instruction in multiple components, this comparison does not identify the cause of their lower coverage.

\subsection{Consistency Across Language Settings}\label{subsec:language}

Table~\ref{tab:main} shows the same observed ordering in the separate Python and C++ evaluations: \tool has the highest OPT@1 and OPT@8 for both backbones. Its OPT@8 margins are 4.98 and 8.35 points in Python, and 12.04 and 12.52 points in C++, for GPT-5-mini and Qwen3-Coder respectively. Because each language uses its own mining corpus and frozen registry, these results demonstrate consistency across two within-language evaluations; they do not show that one library transfers between languages. The larger C++ margins are descriptive; the experiment does not identify their cause. The ordering recurs despite separate corpora and registries.

\begin{table}[!t]
\centering
\small
\begin{tabular}{@{}lrrrr@{}}
\toprule
& \multicolumn{2}{c}{GPT-5-mini} & \multicolumn{2}{c}{Qwen3-Coder} \\
\cmidrule(lr){2-3}\cmidrule(l){4-5}
Variant & OPT@1 & OPT@8 & OPT@1 & OPT@8 \\
\midrule
Full & 26.48 & 37.40 & 21.35 & 36.60 \\
No retrieval & 12.20 & 27.45 & 7.22 & 9.15 \\
Random cards & 13.80 & 27.82 & 12.36 & 13.80 \\
Single plan & 18.78 & 25.84 & 11.88 & 14.29 \\

\bottomrule
\end{tabular}
\caption{Python ablation results for GPT-5-mini and Qwen3-Coder. Full denotes the complete \tool pipeline and the remaining variants modify retrieval relevance or plan diversity.}
\label{tab:ablation}
\end{table}

\subsection{Speedup Magnitude and Failure Analysis}\label{subsec:failure}

\begin{figure}[t]
\centering
\includegraphics[width=\columnwidth]{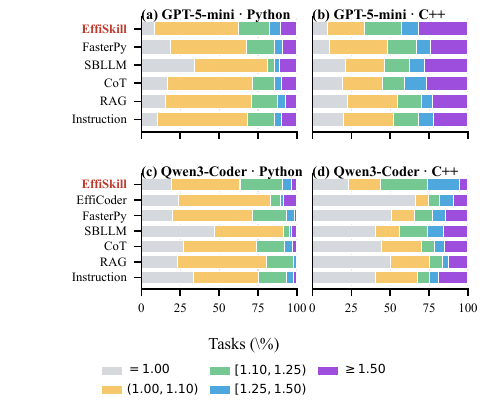}
\caption{Best private-test speedup among the first eight public-ranked candidates. Missing, incorrect, and non-improving outcomes receive the neutral ratio 1.0, separating threshold coverage from improvement magnitude.}
\label{fig:speedup-distribution}
\end{figure}

Figure~\ref{fig:speedup-distribution} complements threshold coverage with improvement magnitude. The clipped all-task geometric means assign missing, incorrect, and non-improving outcomes a neutral ratio of 1.0. \tool obtains 1.188/1.448 for GPT on Python/C++, compared with 1.170/1.432 for the strongest baselines. For Qwen, \tool obtains 1.129/1.176, while the strongest baselines obtain 1.181/1.250. Coverage is \tool's consistent advantage across all four settings; speedup magnitude varies by backbone and language.

Selected private-correct candidates below \(1.10\times\) form the largest non-success category: 60.7\%, 29.4\%, 49.8\%, and 30.7\% of tasks for GPT/Python, GPT/C++, Qwen/Python, and Qwen/C++, respectively. Qwen also has more tasks without a public survivor: 11.7\% versus 1.3\% in Python and 11.1\% versus 2.9\% in C++. Appendix~\ref{app:evaluation} gives the complete terminal count vectors.

\subsection{Analysis of Learned Skills}\label{subsec:skill-analysis}

The Python skill library spans five descriptive families, including implementation-level, algebraic, dynamic-programming, combinatorial, and data-structure transformations. Table~\ref{tab:registry-composition} shows representative mechanisms; Appendix~\ref{app:inventory} gives the complete 29-card inventory, family counts, and card fields.

\begin{table}[t]
\centering
\small
\setlength{\tabcolsep}{4pt}
\begin{tabular}{@{}>{\raggedright\arraybackslash}p{0.32\columnwidth}>{\raggedright\arraybackslash}p{0.61\columnwidth}@{}}
\toprule
Family & Representative transformations \\
\midrule
Implementation \& constants & Constant-factor cleanup (O001); right-sized structures (O006); cheaper arithmetic (O018) \\
Algebraic \& closed-form & Closed-form loops (O009); algebraic counting (O016); digit-range formulas (O024) \\
DP \& state compression & State reformulation (O010); rolling recurrences (O029) \\
Combinatorics \& number theory & Small-\(k\) binomials (O003); coprimality classification (O027) \\
Graphs, sets \& structures & Graph-structure exploitation (O004); set/frequency refactoring (O020) \\
\bottomrule
\end{tabular}
\caption{Representative mechanisms in the released Python Operator-Skill library.}
\label{tab:registry-composition}
\end{table}

\begin{figure}[H]
\centering
\includegraphics[width=\columnwidth]{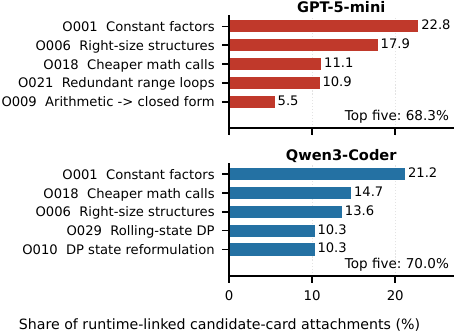}
\caption{Python runtime-linked candidate--card attachment frequencies by backbone. The top five cards account for 68.3\% of GPT-5-mini attachments and 70.0\% of Qwen3-Coder attachments.}
\label{fig:skill-usage}
\end{figure}

Across 16,412 Python attachments, Figure~\ref{fig:skill-usage} shows which cards accompanied runtime-linked candidates under each backbone. O001, O006, and O018 appear among both top-five lists, forming a common implementation-oriented core. The longer tail supplies algorithmic, number-theoretic, graph, and state-specific routes. Attachment frequency characterizes selection by the fixed pipeline and does not isolate a card's causal effect. Thus the shared core and specialized tail play distinct descriptive roles.

\paragraph{Aizu 1464 trace.}
In this example, diagnosis recorded repeated sorting and rational normalization inside roughly 30 parametric-search probes. Retrieval selected O001 and O021. A conservative plan retained the search algorithm while adding buffered input, a local keyed sort, integral probe arithmetic, and one final \texttt{gcd}. Candidate C1 passed pre-execution processing and all public tests, ranked first among public survivors, and passed every private test. Private \texttt{runtime\_sum} was 2,421.0 for the input program and 1,259.6 for the candidate, yielding a \(1.92\times\) speedup.

By preserving the outer parametric search and targeting repeated work inside each probe, the trace illustrates adaptation of a persistent mechanism under task-specific correctness constraints. It exemplifies coordinated local optimization rather than wholesale algorithm replacement.

Appendix~\ref{app:examples} gives the complete evidence chain from diagnosis and retrieved skills through code transformation, public ranking, and private outcomes.

\section{Related Work}\label{sec:related}

Code-efficiency optimization is distinct from code generation and repair: a rewrite must reduce cost while preserving behavior. Benchmarks such as EffiBench, Mercury, EvalPerf, COFFE, and EffiBench-X cover generated code across tasks and languages, while SWE-Perf studies repository-level optimization~\cite{huang2024effibench,du2024mercury,liu2024evaluating,peng2025coffe,qing2025effibench-x,he2025swe}. SWE-Bench ProMax evaluates large-scale multilingual repository refactoring~\cite{shi2026swe}, a complementary setting to execution-efficiency optimization. We optimize existing correct competitive-programming solutions.

Recent optimization methods differ in how they represent and deliver optimization knowledge. \textbf{Prompting} requests rewrites or bottleneck reasoning~\cite{wei2022chain}. \textbf{Retrieval} supplies prior fixes, slow/fast pairs, or summaries~\cite{DBLP:journals/corr/abs-2306-17077,lewis2020retrieval,shypula2023learning,wu2025fasterpy}. \textbf{Learned optimization and fine-tuning} internalize edits or strategies~\cite{DBLP:journals/corr/abs-2208-05297,shypula2023learning,DBLP:journals/corr/abs-2309-14846,huang2024efficoder,yang2025perfcoder}. \textbf{Search and self-evolution} use stochastic search or execution feedback~\cite{petke2017genetic,schkufza2013stochastic,SBLLM,huang2024effilearner,du2025afterburner,hu2026controlled}. These approaches expose knowledge through examples, summaries, learned parameters, or task-specific trajectories. ECO~\cite{kim2025eco} distills and retrieves runtime optimization instructions alongside slow/fast pairs, whereas \tool consolidates recurring transformations into standalone cards used for explicit plan composition.

Research on agent memory and procedural knowledge makes experience available beyond a single prompt. Reflexion retains textual reflections, while Toolformer and ReAct connect language-model reasoning to tools and actions~\cite{shinn2023reflexion,schick2023toolformer,yao2022react}. Packaged-skill interfaces similarly keep reusable instructions outside model parameters~\cite{openai2025codexskills,anthropic2026claude_skills,li2026skillsbench}. \citet{yin2026reasoning} mine recurring reasoning patterns for efficient AI coding, whereas our skills describe program-transformation procedures. Operator Skills specialize this perspective to performance transformations: they preserve an applicability-conditioned procedure rather than a concrete demonstration or task-bound history.

Complementary work targets context pruning~\cite{wang2026swe}, parallel reasoning~\cite{zhang2026paratempo}, early prediction of agent-evaluation outcomes~\cite{shi2026earlyeval}, and repository-setup latency~\cite{shi2026speculative}; these objectives concern agent execution rather than the runtime of generated programs.

Planning provides the adaptation layer between persistent knowledge and a new program. A generated plan is transient and task-specific; it can select, combine, order, or reinterpret mechanisms in response to a diagnosis~\cite{wei2022chain,yao2022react}. In contrast, an Operator Skill remains stable and separately retrievable. \tool therefore separates reusable optimization mechanisms from the plans and candidate trajectories that apply them. Cards persist across tasks, whereas diagnoses and plans are regenerated; unlike parameterized knowledge, their procedures remain inspectable. The experiment, however, does not isolate this representation from retrieval and planning.

\section{Discussion}\label{sec:discussion}

\tool shifts the level at which optimization experience is reused. Prompting and exemplar retrieval primarily expose instance-level code, whereas Operator Skills externalize recurring slow-to-fast transformations as explicit mechanisms. As a result, the reusable unit is a transformation mechanism rather than an instance. Diagnosis can therefore connect a task-specific bottleneck to a procedure even when the new program differs syntactically from stored examples. The Aizu trace illustrates this distinction, but the aggregate experiments do not isolate cards from all other pipeline changes.

A second factor is compositionality. Many performance issues require coordinated changes across multiple aspects of a program, such as state representation, transition structure, and constant-factor optimization. A persistent card defines a reusable mechanism; a task-specific plan determines how one or more mechanisms should be combined, ordered, and realized under the current program's constraints. Multiple plans explore coherent alternatives without changing the frozen knowledge source. This structured exploration constrains the search space while preserving diversity. The matched ablation associates this diversity with higher coverage than a single-plan variant, while not establishing which individual plan or card caused a success.

Third, the skill library supplies an explicit structured prior over generation. Conditioning on a small set of bottleneck-relevant mechanisms narrows attention toward targeted rewrites and makes the supplied optimization knowledge inspectable. This is particularly important in our setting, where optimization is performed without execution-time feedback. This mechanism-level prior effectively regularizes generation, which may help explain why \tool consistently outperforms direct prompting under the same backbone model. The retrieval ablations support the importance of relevant procedural context at the system level, but a direct comparison against raw pairs, traces, or unstructured summaries remains necessary to attribute the gain specifically to the card representation.

Finally, the failure profile separates two opportunities. The largest non-success category consists of private-correct candidates below the speed threshold, suggesting a need for stronger transformations rather than correctness repair alone. Qwen also produces more tasks without a public survivor, highlighting candidate reliability. These terminal outcomes do not support assigning failures to retrieval, planning, compilation, or generation stages.

\section{Conclusion}\label{sec:conclusion}

We presented \tool, a two-stage code-efficiency framework that distills recurring slow-to-fast transformations as Operator Skills. Separating offline mining from fixed skill-guided generation enables mechanism-level reuse beyond one-shot rewriting or instance-level exemplar matching.

Across 623 EffiBench-X tasks per language, \tool attains the highest observed OPT@1 and OPT@8 among the compared methods in all four settings, with OPT@8 margins of 4.98--12.52 percentage points. Matched ablations link relevant retrieval and multiple plans to broader coverage; speedup magnitude varies by setting.

\section*{Limitations}\label{sec:limitations}

This study evaluates competitive-programming tasks in Python and C++ using two model backbones; extending the analysis to additional languages, models, and repository-scale settings remains future work. All methods share the same maximum output budget, although their total inference costs are not matched. Finally, our ablations assess retrieval relevance and plan diversity within the complete framework, but do not fully disentangle the contribution of the Operator Skill representation from the surrounding retrieval and planning pipeline.

\section*{Ethical Considerations}\label{sec:ethics}

This study uses competitive-programming benchmarks and does not involve human participants or personal data. \tool may reduce software execution costs, but automatically generated optimizations can silently alter behavior or introduce security flaws; outputs should therefore be reviewed and tested before deployment. Skill mining and LLM inference also consume computational resources, so practical use should weigh these costs against downstream efficiency gains.

\begingroup
\small
\bibliography{effiskill}
\endgroup

\appendix
\raggedbottom
\section{Frozen Operator-Skill Inventory}
\label{app:inventory}

This appendix presents the 29 Python Operator Skills used in the library
analysis.  Each card records an optimization mechanism, applicability signals,
an expected effect, and correctness risks.  Four fixed controller resources
S1--S4 guide diagnosis, retrieval, planning, and generation in Stage~II; they
are orchestration prompts rather than mined Operator Skills.

The five groups below form the descriptive mapping used throughout the paper.
They contain 5, 8, 3, 10, and 3 cards, respectively, and cover O001--O029
exactly once.  Each entry states the trigger and expected transformation at a
level that can be applied to a new program.  Applicability always depends on
the target program's constraints and interface.

\subsection{Implementation and constant-factor transformations (5)}

\paragraph{O001: Python Constant Factor Optimization and Lightweight Data
Structure Refactor.}
\textbf{Trigger/description:} an asymptotically appropriate Python solution
still approaches the time limit, especially when templates, wrapper calls,
heavy containers, repeated allocation, or hot-loop I/O dominate.
\textbf{Expected effect:} strip unused infrastructure, tighten hot work, and
replace general structures with lighter task-specific ones, normally reducing
runtime and allocation without changing the governing \(O(n)\) or
\(O(n\log n)\) complexity.

\paragraph{O006: Right Size Structures and Stream the Work.}
\textbf{Trigger/description:} fixed-size buffers, full-range scans, redundant
passes, or dense structures scale with a magic bound rather than the observed
input.
\textbf{Expected effect:} stream aggregates and right-size boolean,
frequency, or hash structures so time and memory track actual data while
removing hidden \(O(K)\) work and avoidable allocations.

\paragraph{O011: Optimize Per Digit Loops in Brute Force Number Ranges.}
\textbf{Trigger/description:} a large range loop performs per-number digit
work through Python-level modulo/division loops, helper calls, float
arithmetic, or unnecessary tables.
\textbf{Expected effect:} flatten the hot path and use safe integer or
C-implemented operations to lower per-number overhead; this is a
constant-factor transformation and does not replace the whole range scan.

\paragraph{O018: Replace Heavy Math \& Library Calls with Cheap Arithmetic or
Log Based Forms.}
\textbf{Trigger/description:} tight loops repeatedly invoke
\texttt{log}/\texttt{log2}, \texttt{ceil}, floating-point exponentiation, or
other scalar library operations.
\textbf{Expected effect:} simplify the algebra, maintain reusable products,
or use integer/bit forms and direct arithmetic to remove expensive calls from
the hot loop, generally without claiming an asymptotic change.

\paragraph{O028: Streaming Single Pass Product With Overflow Guard.}
\textbf{Trigger/description:} code computes a product over many inputs and
separately scans for zero or a fixed overflow/threshold condition.
\textbf{Expected effect:} fold product, zero handling, and overflow checks
into one pass with early termination and \(O(1)\) auxiliary state.

\subsection{Algebraic and closed-form reformulations (8)}

\paragraph{O005: Mean Based Minimization of Sum of Squared Deviations.}
\textbf{Trigger/description:} every integer center between the data minimum
and maximum is tried, and the squared-error sum is recomputed for each center.
\textbf{Expected effect:} evaluate the integer candidates around the mean,
optionally using the sum and sum of squares, replacing range-dependent
repeated scans with an \(O(n)\) aggregation.

\paragraph{O009: Arithmetic Loop to Closed Form Optimization.}
\textbf{Trigger/description:} arithmetic progressions, floor-division
patterns, divisor sums, or simple numeric loops are evaluated by direct
iteration.
\textbf{Expected effect:} use a formula, quotient/divisor grouping, prefix
aggregate, or bounded enumeration to reduce heavy loops to \(O(1)\),
\(O(\sqrt{N})\), or another input-appropriate bound.

\paragraph{O012: Mean Based Fixed-Range Sum of Squared Deviations
Minimization.}
\textbf{Trigger/description:} an outer loop over a fixed integer range
repeatedly scans all values to compute
\(\sum_i(a_i-p)^2\).
\textbf{Expected effect:} choose the mean-derived integer center and compute
the objective once, collapsing repeated quadratic-cost scans to a single
aggregation.

\paragraph{O016: Eliminate Bruteforce Triples via Algebraic Counting.}
\textbf{Trigger/description:} three integer loops perform simple arithmetic
under a linear equality, range, or ordering constraint, and only the count or
existence of solutions is needed.
\textbf{Expected effect:} solve one variable, tighten the remaining bounds,
or count feasible values directly, typically reducing \(O(n^3)\) enumeration
to \(O(n^2)\), \(O(n)\), or a closed form where the constraints permit it.

\paragraph{O021: Refactor Redundant Range Loops and Pairwise Aggregations.}
\textbf{Trigger/description:} a loop repeatedly recomputes an almost-global
\texttt{sum}, extremum, range scan, or pairwise aggregate.
\textbf{Expected effect:} precompute aggregates and use prefix identities,
pairwise identities, convexity, or a small sign enumeration to remove
redundant quadratic or range-dependent work.

\paragraph{O022: Algebraic Loop Elimination for Integer Combinatorics.}
\textbf{Trigger/description:} nested integer loops share a size parameter and
the innermost body enforces only simple linear arithmetic and comparisons.
\textbf{Expected effect:} make dependent variables explicit and tighten
monotone bounds, eliminating one or more enumeration dimensions while
preserving integrality, domain, and ordering checks.

\paragraph{O024: Digit Range Closed Form Counting.}
\textbf{Trigger/description:} a loop from 1 to \(N\) computes a digit-length
or magnitude-bucket property independently for every integer.
\textbf{Expected effect:} sum over ranges with equal digit length, replacing
the full scan with a piecewise arithmetic count.

\paragraph{O025: Replace Enumeration and Simulation with Parity/Bitwise
Closed Forms.}
\textbf{Trigger/description:} combinatorial objects or simple numeric
processes are explicitly enumerated even though only parity, bitwise
structure, or a small aggregate state affects the answer.
\textbf{Expected effect:} derive a parity/bitwise/counting formula or direct
construction, removing the enumeration or simulation when its assumptions
hold.

\subsection{DP and state-compression transformations (3)}

\paragraph{O010: DP State Reformulation and Dimension/Data Structure
Optimization.}
\textbf{Trigger/description:} a dynamic program stores an item/step dimension
whose transitions use only current or previous state, or scans all prior
states for each transition.
\textbf{Expected effect:} collapse redundant dimensions, prune unreachable
states, and use an appropriate range structure when needed, reducing memory
and transition cost.

\paragraph{O013: Incremental State Exponential Search Refactor.}
\textbf{Trigger/description:} an inherent exponential search reconstructs
lists, sums, counters, or merged state at every node.
\textbf{Expected effect:} represent state with scalars or bitmasks and update
aggregates incrementally; the inherent exponential state count may remain,
but per-state work and allocation fall.

\paragraph{O029: Rolling State Linear DP \& Recurrence Compression.}
\textbf{Trigger/description:} position-indexed DP transitions depend only on
a fixed window of preceding states.
\textbf{Expected effect:} roll the recurrence into \(O(1)\) or fixed-size
state, stream inputs, and simplify fixed transitions instead of retaining the
full history.

\subsection{Combinatorics and number-theory transformations (10)}

\paragraph{O003: Small k Binomial via Prime Exponents (No Global
Factorials).}
\textbf{Trigger/description:} global factorial and inverse-factorial tables
are built although only a few binomials with small complementary parameter
are needed.
\textbf{Expected effect:} use factor exponents or a direct small-\(k\)
multiplicative computation, removing \(O(N)\) tables and scaling work to the
small parameter.

\paragraph{O007: Input Scaled Binomial Mod Optimization.}
\textbf{Trigger/description:} factorial, inverse, or Pascal tables are sized
to a fixed worst-case constant for one or very few modular-combination
queries.
\textbf{Expected effect:} size any setup to the observed input or use an
incremental/multiplicative combination, lowering avoidable setup time and
memory.

\paragraph{O008: Input sensitive coprimality and prime structure
classification.}
\textbf{Trigger/description:} an integer collection must be classified by
shared prime factors, but code uses pairwise checks, heavy per-element
factorization, or a sieve wider than the observed values.
\textbf{Expected effect:} sieve only to the actual maximum, track used primes
with lightweight arrays or frequencies, and early-exit when the classification
is determined.

\paragraph{O014: Right size or Remove Heavy Combinatorial Precomputation.}
\textbf{Trigger/description:} large fixed factorial/inverse tables or heavy
numeric libraries are initialized independently of the actual combination
arguments.
\textbf{Expected effect:} use input-sized pure-Python tables or direct
small-\(k\) multiplication so setup and memory match the queries.

\paragraph{O015: Optimize Modular Combinatorics And Binomials.}
\textbf{Trigger/description:} a modular \(n\)-choose-\(k\) helper recomputes
inverses, powers, or closely related terms inside every iteration.
\textbf{Expected effect:} use complementary small \(k\), running/reusable
powers and inverses, and algebraic simplification to reduce modular-loop
overhead.

\paragraph{O017: Factorial Divisor Prime Exponent Aggregation Optimization.}
\textbf{Trigger/description:} a divisor count or related quantity needs the
prime exponents of \(n!\) or another consecutive product, but integers are
refactored repeatedly.
\textbf{Expected effect:} perform one sieve and aggregate exponents with
Legendre-style sums or an equivalent global pass, avoiding repeated
factorization.

\paragraph{O019: Divisor Enumeration and GCD Folding Optimization.}
\textbf{Trigger/description:} loops proportional to a raw numeric value do
only modulo/divisibility work, or an array repeatedly computes common-divisor
relationships.
\textbf{Expected effect:} enumerate paired divisors in \(O(\sqrt{V})\) or
fold Euclidean GCD in \(O(n\log V)\), replacing raw-value scans when the
algebra supports it.

\paragraph{O023: Modular Binomial Combinatorics Without Big Int Factorials.}
\textbf{Trigger/description:} exact large factorials and integer division are
used to form a binomial before taking a modulus.
\textbf{Expected effect:} use modular factorial/inverse tables or a
small-parameter multiplicative formula with early reduction, preventing
arbitrary-precision factorial growth.

\paragraph{O026: Control Big Integer Growth in Factorization Based
Combinatorics.}
\textbf{Trigger/description:} factorization-derived combinatorial expressions
construct large factorials or products before modular reduction.
\textbf{Expected effect:} use exponent reasoning, multiplicative or modular
combinations, and early reduction to keep intermediates bounded.

\paragraph{O027: Coprimality Classification Optimization.}
\textbf{Trigger/description:} pairwise GCD or per-number trial division is
used to distinguish pairwise from setwise coprimality.
\textbf{Expected effect:} select a maximum-aware sieve/SPF or
frequency--divisor strategy, share preprocessing, and stop when repeated prime
use is detected.

\subsection{Graph, data-structure, and set transformations (3)}

\paragraph{O002: Bfs Constructive Kth Digit Sequence.}
\textbf{Trigger/description:} the \(k\)-th value of a sequence defined by
local digit constraints is obtained by over-generating, sorting, recursive
enumeration, or binary-searching a digit DP.
\textbf{Expected effect:} breadth-first generation from ordered seeds with
ordered children stops after \(k\) outputs, using \(O(kB)\) time and \(O(k)\)
queue space when FIFO order matches the required numeric or lexicographic
order.

\paragraph{O004: Exploit Graph Structure and Data Structures for Faster
Python Graph Algorithms.}
\textbf{Trigger/description:} a sparse or highly structured graph is processed
with repeated traversals, dense matrices, object-heavy queues, or generic
connectivity/flow machinery.
\textbf{Expected effect:} exploit structure and use adjacency lists,
lightweight BFS/DFS, DSU, or a problem-sized flow representation to remove
dense or repeated work; the exact bound depends on the graph structure.

\paragraph{O020: Set Intersection or Frequency Array Refactor for Common
Elements Queries.}
\textbf{Trigger/description:} nested group/item loops repeatedly perform
linear list membership or removal to find values common to several groups.
\textbf{Expected effect:} intersect hash sets or aggregate bounded
frequencies once, replacing repeated membership scans with work proportional
to total input size.

\subsection{Expanded O016 operator specification}
\label{app:o016-full}

O016 targets counting problems in which three bounded integer variables are
linked by a linear equality.  If one variable can be derived from the other
two, explicit cubic enumeration is unnecessary.

\paragraph{Applicability.}
Use O016 when the program counts solutions rather than listing them, the
innermost variable participates only in a linear relation and range or ordering
checks, and the remaining predicates can be translated into exact bounds.

\paragraph{Procedure.}
Choose two independent variables, solve the equality for the third, and derive
its feasible interval from the original domain and ordering constraints.  If
every value in an interval is valid, add the interval length directly; if an
integrality condition remains, count only values in the required residue
class.  Tighten the outer bounds before iteration.

\paragraph{Expected effect.}
Direct elimination reduces a typical $O(n^3)$ count to $O(n^2)$ with
$O(1)$ auxiliary space.  When one remaining dimension can also be counted in
closed form, the time can fall to $O(n)$ or $O(1)$.

\paragraph{Correctness checks.}
The derived variable must be integral and inside its original domain.  Strict
ordering and distinctness constraints must remain strict after substitution,
and inclusive interval endpoints must be checked for off-by-one errors.  The
operator does not apply when solutions must be materialized or when nonlinear,
data-dependent predicates prevent exact bound derivation.

\paragraph{Minimal example.}
For $1\leq a,b,c\leq n$ and $a+b+c=S$, fixing $a$ leaves
$c=S-a-b$.  Valid $b$ values lie in

\[
\max(1,S-a-n)\leq b\leq\min(n,S-a-1),
\]

so one interval-length calculation replaces the inner two loops:

\begin{Verbatim}[breaklines=true,fontsize=\small]
count = 0
for a in range(1, n + 1):
    lo = max(1, S - a - n)
    hi = min(n, S - a - 1)
    if lo <= hi:
        count += hi - lo + 1
\end{Verbatim}

\section{Stage I: Mining and Registry Construction}
\label{app:stage1}

Stage~I converts paired slow and fast programs into a frozen, language-specific
library of reusable Operator Skills.  The pipeline contains five steps:
canonicalize pairs, extract structured traces, form deterministic signatures,
cluster related mechanisms, and consolidate cluster summaries into cards.

\subsection{Pair normalization}

For each language, the mining corpus contains 1,800 directional rows: 900
slow-to-fast comparisons and their 900 reversed counterparts.  Direction
normalization restores the canonical slow/fast order, and source-hash
deduplication leaves 900 unique pairs per language before trace extraction.

The C++ pairs come from PIE~\cite{shypula2023learning}.  The Python pairs are
drawn from PIE, Mercury~\cite{du2024mercury}, DeepMind Code
Contests~\cite{li2022competition}, and the CodeParrot distribution of
APPS~\cite{hendrycks2021apps}.  These mining sources are used only in Stage~I;
the EffiBench-X~\cite{qing2025effibench-x} evaluation tasks are kept separate
and do not contribute to registry construction.

\paragraph{Terms and redistribution.}
At their official distribution points, EffiBench-X is Apache-2.0, Mercury is
CC BY-NC 4.0, DeepMind Code Contests places code under Apache-2.0 and non-code
material under CC BY 4.0, CodeParrot APPS is MIT, and Qwen3-Coder is
Apache-2.0~\cite{qwen30b}.  This paper does not redistribute raw third-party
tasks, tests, programs, mining rows, or model weights.

The adapter accepts explicit \texttt{slow\_code}/\texttt{fast\_code} rows,
labeled A/B comparisons, and input/output rows with a direction field.  It
removes surrounding code fences, trims text, and serializes the following
canonical record:

\begin{Verbatim}[breaklines=true,breakanywhere=true,fontsize=\scriptsize]
{
  "problem_id": <id>,
  "statement": <normalized statement>,
  "slow_code": <normalized slow source>,
  "fast_code": <normalized fast source>,
  "speedup_ratio": <optional numeric value>
}
\end{Verbatim}

Empty programs are rejected.  Deduplication operates on the canonicalized
record rather than approximate source similarity, so reversing a pair does not
create a second optimization experience.

\subsection{Trace extraction}

GPT-5.1 converts each canonical pair into a structured optimization trace with
four components:

\begin{itemize}
    \item \texttt{ProblemBrief}: inferred constraints and problem-type tags;
    \item \texttt{SlowAudit}: dominant operations, complexity, and typed
          bottlenecks;
    \item \texttt{FastAudit}: the faster program's core mechanism and
          complexity; and
    \item \texttt{DeltaSummary}: transformation type, before/after complexity,
          steps, trigger signals, and pitfalls.
\end{itemize}

The extraction instruction requires JSON only.  The user-template contract is:

\begin{Verbatim}[breaklines=true,breakanywhere=true,fontsize=\scriptsize]
Return JSON only with keys:
problem_id,
ProblemBrief{constraints_guess{n_max,q_max,notes},problem_type_tags},
SlowAudit{dominant_ops,complexity{time,space,why},
          bottlenecks[{kind,detail}]},
FastAudit{core_idea,complexity{time,space,why}},
DeltaSummary{delta_type,complexity_delta{from,to,note},
             transformation_steps,trigger_signals,pitfalls}.

problem_id: <problem_id>
statement: <statement>
slow_code: <slow_code>
fast_code: <fast_code>
\end{Verbatim}

The parser first reads strict JSON and then, if necessary, extracts the
outermost JSON object and requests one format-only repair.  Schema validation
normalizes list and scalar fields before downstream processing.

\subsection{Deterministic signatures}

Trace \(i\) receives identifier \texttt{T\%07d}.  Its textual signature
concatenates, in order:

\begin{enumerate}
    \item normalized \texttt{delta\_type} values;
    \item the before-to-after complexity change;
    \item up to eight trigger signals;
    \item up to eight transformation steps;
    \item typed slow-code bottleneck kinds; and
    \item up to six problem-type tags.
\end{enumerate}

Field separators preserve component boundaries.  Problem identifiers,
statement wording, and output formatting are excluded, concentrating the
representation on the computational transformation.

\subsection{Hybrid clustering}

The Stage-I clustering implementation combines lexical and semantic views of
each signature.  Table~\ref{tab:stage1-clustering} documents its configuration.

\begin{table}[H]
\centering
\small
\begin{tabular}{@{}p{.32\columnwidth}p{.61\columnwidth}@{}}
\toprule
Component & Configuration \\
\midrule
Lexical view & word TF--IDF, unigrams and bigrams,
\texttt{max\_features=4096}, row-normalized \\
Semantic view & normalized
\texttt{all-MiniLM-L6-v2} embeddings \\
Fallback & TF--IDF truncated SVD, at most 128 dimensions, seed 42 \\
Fusion & concatenation of normalized lexical and semantic views \\
Candidate (k) & even values in a clipped \(\pm10\) window around
\(\mathrm{round}(\sqrt n)\), bounded by 6 and 80 \\
Selection & KMeans with seed 42; maximum cosine silhouette \\
\bottomrule
\end{tabular}
\caption{Stage-I hybrid clustering configuration.}
\label{tab:stage1-clustering}
\end{table}

If semantic embeddings are not available, the SVD view supplies the dense
component.  Row normalization prevents either view from dominating solely
because of vector scale.

\subsection{Skill construction and consolidation}

For each provisional cluster, GPT-5.1 receives aggregate
transformation types, complexity changes, bottlenecks, tags, triggers, steps,
support, and representative traces.  It returns one primary mechanism together
with a name, description, applicability signals, expected effect, procedure,
pitfalls, complexity, and exclusions.  Problem and contest names are omitted
to keep the result reusable across tasks.

Programmatic post-processing normalizes names and bullet lists, bounds field
lengths, validates categorical values, and compares descriptions, tags,
triggers, steps, and family information.  Consolidation computes word TF--IDF
unigram/bigram similarity with \texttt{max\_features=2048}.  Two summaries can
merge only when their scopes are compatible and either their normalized names
match or their cosine similarity reaches 0.76 while trigger Jaccard overlap
(or tag overlap when triggers are absent) is at least 0.20.  Connected
components define merge groups, after which the consolidated records are
rendered as stable Operator Skill cards.  These values define the consolidation
configuration.

\subsection{Complete Stage-I procedure}

\begin{algorithm}[H]
\caption{Stage-I Operator Skill construction}
\label{alg:stage1-full}
\begin{algorithmic}[1]
\REQUIRE directional rows $D$, configuration $\theta$
\STATE $U\leftarrow[\,]$, $H\leftarrow\emptyset$
\FOR{row $r\in D$}
  \STATE $p\leftarrow\mathrm{OrientNormalize}(r)$
  \STATE $h\leftarrow\mathrm{SHA256}(\mathrm{CanonicalJSON}(p))$
  \IF{\(h\notin H\)}
    \STATE append $(p,h)$ to $U$; $H\leftarrow H\cup\{h\}$
  \ENDIF
\ENDFOR
\FOR{\((p,h)\in U\)}
  \STATE $t\leftarrow\mathrm{ExtractTrace}_{\text{GPT-5.1}}(p;\theta_{\rm trace})$
  \STATE $s\leftarrow\mathrm{TraceSignature}(t)$
\ENDFOR
\STATE $X_{\rm lex}\leftarrow
  \mathrm{Normalize}(\mathrm{TFIDF}_{1,2}(s))$
\STATE $X_{\rm sem}\leftarrow\mathrm{Normalize}(\mathrm{Embed}(s))$
\STATE $X\leftarrow[X_{\rm lex};X_{\rm sem}]$
\STATE $k\leftarrow\mathrm{SelectCosineSilhouette}(X)$
\STATE $C\leftarrow\mathrm{KMeans}(X,k;\mathrm{seed}=42)$
\FOR{cluster $c\in C$}
  \STATE $g_c\leftarrow\mathrm{AggregateProfile}(c)$
  \STATE $a_c\leftarrow\mathrm{SummarizeValidate}(g_c;\theta_{\rm summary})$
\ENDFOR
\STATE $A\leftarrow\mathrm{MergeCompatible}(\{a_c\})$
\STATE \textbf{return} frozen registry \(\mathrm{RenderSerialize}(A)\)
\end{algorithmic}
\end{algorithm}

The result is a language-specific registry whose cards expose the reusable
mechanism independently of any one source program.  Appendix~\ref{app:inventory}
lists the complete Python Operator Skill inventory used in the library
analysis.  The pseudocode documents the deterministic construction procedure;
the frozen registry is used unchanged in Stage~II.

\FloatBarrier

\section{Stage II: Schemas, Routing, and Candidate Handling}
\label{app:stage2}

Stage~II applies a frozen within-language registry to a previously unseen
program.  Four fixed controller resources, S1--S4, define the interfaces for
diagnosis, retrieval, planning, and generation.  They orchestrate the Operator
Skills but are not themselves mined optimization knowledge.

\subsection{Data contracts}

The controller passes four typed records between stages.  Angle-bracketed
values below are supplied by the task or the selected skill set.

\begin{Verbatim}[breaklines=true,breakanywhere=true,fontsize=\scriptsize]
S1 OptimizationBrief
  constraints_guess: {n_max, q_max, notes}
  problem_type_tags: [string]
  dominant_ops: [string]
  complexity: {time, space, why}
  bottlenecks: [{kind, detail}]
  must_change: algorithm | implementation | mixed
  risks: [string]
  confidence: high | medium | low

S2 SkillSets
  skill_sets: [{
    set_id,
    route_type: algorithmic_rewrite | data_structure |
                memory_layout | io | constant_factor |
                math_precompute | fallback,
    confidence, selection_theme, expected_change,
    selected_skills: [{skill_id, reason, expected_role}]
  }]
  global_coverage_gaps: [string]

S3 Plans
  plans: [{
    plan_id, route_type, plan_style,
    name, core_idea, target_complexity,
    optimization_mechanism, old_hot_path, new_mechanism,
    expected_speedup_band, impact_skill_ids, risks, steps
  }]

S4 Candidate
  code: string
  complexity: {time, space, why}
  dominant_ops: [string]
  change_summary: {removed_hot_path, new_mechanism,
                   expected_effect}
  static_verdict: pass | fail
  interface_match: boolean
  complexity_plausible: boolean
  static_issues: [string]
  static_confidence: number in [0,1]
\end{Verbatim}

The S1 message contains the problem statement and baseline program.  S2
receives the diagnosis and registry metadata.  S3 receives the task, baseline,
diagnosis, one selected mechanism set, and the complete bodies of its selected
cards.  S4 receives the task, baseline, and one plan.  Every candidate must
preserve the original callable or script interface and visibly implement the
plan's named mechanism.

\subsection{Controller instructions}

\paragraph{Diagnosis (S1).}
The diagnostician identifies inferred constraints, dominant operations,
baseline complexity, likely bottlenecks, required change type, and semantic
risks.  It explicitly checks for unused computation, avoidable sorting,
materialization, repeated traversal, and unnecessarily general data
structures.

\paragraph{Retrieval (S2).}
The router selects one or more distinct mechanism sets using only valid IDs
from the frozen registry.  Each selected card must match a diagnosed
bottleneck, and each set states the intended code- or complexity-level change.
Different sets are encouraged to cover genuinely different optimization
routes rather than paraphrases of one idea.

\paragraph{Planning (S3).}
For every mechanism set, the planner adapts the selected cards to the target
program.  A plan names the old hot path, the replacement mechanism, the target
complexity, the ordered implementation steps, relevant skill IDs, and
correctness risks.  Plans must preserve the external interface and differ in a
substantive runtime-reduction mechanism or in a meaningful application of the
same mechanism.

\paragraph{Generation (S4).}
The generator emits one complete implementation for an allocated plan together
with a complexity explanation and change summary.  Cosmetic rewrites are
rejected by the contract: the requested hot-path change must be present in the
code.

\subsection{Parsing and normalization}

Each stage requests strict JSON.  The parser removes a surrounding code fence,
attempts strict parsing, and can extract the outermost object and request one
format-only repair.  It then normalizes route names, list/scalar fields, skill
IDs, and confidence values.  Unknown IDs and empty mechanism sets are removed
before planning; an empty code object never enters the candidate pool.

\subsection{Plan diversity and slot allocation}

The router produces at most three mechanism sets with at most six selected
Operator Skills per set.  The planner can produce conservative, balanced, and
aggressive applications, together with a whole-program simplification route.
This creates more possible plans than the output budget can accommodate.

Slots are therefore assigned before any candidate execution.  The allocator
scores each plan by expected impact and diagnosis alignment, then penalizes
repeated routes, mechanism sets, and normalized mechanisms.  It first covers
distinct high-upside plans and then distributes remaining slots in score
order.  The allocations sum to at most (K=8).  Failed or unparsable
generations leave an unfilled slot; they are not replaced using execution
feedback.

\subsection{Pre-execution processing}

Candidate code is cleaned without changing its internal structure.  Python
processing parses the abstract syntax tree and compares class, function,
method, and script interfaces.  C++ processing records parser- and
compiler-facing diagnostics through its language-specific path.  These checks
occur after generation and before benchmark execution.  They do not use public
or private test outcomes.

\subsection{Complete Stage-II procedure}

\begin{algorithm}[H]
\caption{Stage-II skill-guided optimization}
\label{alg:stage2-full}
\begin{algorithmic}[1]
\REQUIRE task context $C$, program $S$, registry $\mathcal S$, bound $K$
\STATE $b\leftarrow\mathrm{Diagnose}(C,S)$
\STATE $M\leftarrow\mathrm{Metadata}(\mathcal S)$
\STATE \(\mathcal R\leftarrow\mathrm{NormalizeSets}(\mathrm{Route}(b,M))\)
\STATE \(\mathcal P\leftarrow[\mathrm{WholeProgramSimplification}(b)]\)
\FOR{set $r\in\mathcal R$}
  \STATE $B_r\leftarrow\mathrm{FullCardBodies}(r,\mathcal S)$
  \STATE \(\mathcal P\leftarrow\mathcal P\mathbin{+\!\!+}
    \mathrm{NormalizePlans}(\mathrm{Plan}(C,S,b,r,B_r))\)
\ENDFOR
\STATE $a\leftarrow\mathrm{AllocateSlots}(\mathcal P,K,b)$
\STATE \(\mathcal O\leftarrow[\,]\)
\FOR{plan $p\in\mathcal P$}
  \FOR{slot $j=1,\ldots,a_p$}
    \STATE $o\leftarrow\mathrm{Generate}(C,S,p)$
    \IF{$o$ contains nonempty code}
      \STATE attach plan, route, skill, and local diagnostic metadata
      \STATE append $o$ to \(\mathcal O\)
    \ENDIF
  \ENDFOR
\ENDFOR
\STATE serialize \(\mathcal R,\mathcal P,\mathcal O\) before execution
\STATE \textbf{return} \(\mathcal O\) and its diagnostics
\end{algorithmic}
\end{algorithm}

\subsection{Execution and reporting boundary}

Generation and pre-execution processing finish before any candidate is run.
Public tests then execute candidates independently, remove incorrect outputs,
and order survivors by aggregate public \texttt{runtime\_sum}.  The first
$k$ entries of this single order define the OPT@$k$ prefix.  Held-out
private tests determine reported correctness and speedup but do not alter
generation, retrieval, plans, allocation, public filtering, or public order.
Empty or short pools remain failures at missing positions and never change the
623-task denominator.

\FloatBarrier

\section{Evaluation and Statistical Detail}
\label{app:evaluation}

\begingroup
\setcounter{dbltopnumber}{1}
\renewcommand{\dbltopfraction}{.80}
\renewcommand{\dblfloatpagefraction}{.95}

\subsection{Task split and evaluation order}

EffiBench-X contributes 623 tasks and 100 tests per task in each evaluated
language.  A fixed per-task partition assigns 20 cases to public selection and
80 to held-out private evaluation, giving 12,460 public and 49,840 private test
cases per language.

For every method, the canonical input program is evaluated against the same
task tests.  Candidate generation and language-specific pre-execution
processing finish first.  Public correctness filters the pool, aggregate
public \texttt{runtime\_sum} orders survivors, and the private records at the
same task/output coordinates determine the final outcome.  Private data never
select or reorder candidates.  The formal task and aggregate definitions
appear once, in Section~\ref{subsec:protocol}.

\subsection{Complete paired-bootstrap results}

Each endpoint compares \tool with the strongest observed non-\tool baseline
for that endpoint.  The unit of resampling is the task-aligned binary success
difference.  We use 10,000 one-sided paired bootstrap replicates, report the
2.5th and 97.5th percentiles of the percentage-point mean difference, and
apply Holm's step-down adjustment jointly to the eight raw \(p\)-values.
Explicit endpoint seeds make the resamples independent of Python's randomized
\texttt{hash()}.

The sole nonsignificant comparison is GPT/C++ OPT@1.  No raw-\(p\) marker is
used in the main table.

\begin{table*}[t]
\centering
\scriptsize
\begin{tabular}{@{}lllrrrrrl@{}}
\toprule
Backbone & Language & Endpoint / baseline & Margin & 95\% CI & Raw \(p\) &
Holm \(p\) & Seed & \(<.05\) \\
\midrule
GPT-5-mini & Python & OPT@1 / FasterPy & 5.297 & [1.766, 8.828] & .001500 & .007499 & 3442445853 & yes \\
GPT-5-mini & Python & OPT@8 / FasterPy & 4.976 & [0.642, 9.310] & .013899 & .038396 & 1694924378 & yes \\
GPT-5-mini & C++ & OPT@1 / CoT & 3.692 & [$-$0.963, 8.507] & .070593 & .070593 & 3503316074 & no \\
GPT-5-mini & C++ & OPT@8 / CoT & 12.039 & [7.544, 16.533] & .000100 & .000800 & 2198235361 & yes \\
Qwen3-Coder & Python & OPT@1 / CoT & 5.297 & [1.284, 9.149] & .004200 & .016798 & 1800306935 & yes \\
Qwen3-Coder & Python & OPT@8 / FasterPy & 8.347 & [3.692, 13.002] & .000400 & .002400 & 495226222 & yes \\
Qwen3-Coder & C++ & OPT@1 / SBLLM & 5.778 & [0.803, 10.754] & .012799 & .038396 & 1186028528 & yes \\
Qwen3-Coder & C++ & OPT@8 / SBLLM & 12.520 & [7.544, 17.496] & .000100 & .000800 & 2590448128 & yes \\
\bottomrule
\end{tabular}
\caption{Complete deterministic inference table.  Margins and intervals are
percentage points.  Main-table stars encode the final Holm decision only.}
\label{tab:full-statistics}
\end{table*}

\subsection{Candidate-prefix values}

Table~\ref{tab:all-topk-values} contains all 208 plotted values: every
applicable method, setting, and \(k=1,\ldots,8\).
Every method is evaluated under the same at-most-eight-output reporting bound.
If generation or parsing leaves fewer than eight usable outputs, missing
positions contribute no success and the task remains in the denominator.
\begin{table*}[t]
\centering
\scriptsize
\renewcommand{\arraystretch}{0.84}
\setlength{\tabcolsep}{2.7pt}
\begin{tabular}{@{}llrrrrrrrr@{}}
\toprule
Setting & Method & OPT@1 & OPT@2 & OPT@3 & OPT@4 & OPT@5 & OPT@6 & OPT@7 & OPT@8 \\
\midrule
GPT/Python & Instruction & 18.62 & 21.83 & 24.88 & 26.97 & 27.93 & 30.34 & 31.30 & 31.62 \\
 & RAG & 19.90 & 23.43 & 25.52 & 27.29 & 27.93 & 28.41 & 29.21 & 29.37 \\
 & CoT & 18.62 & 22.31 & 24.24 & 25.36 & 26.65 & 27.77 & 28.25 & 28.73 \\
 & SBLLM & 17.34 & 17.82 & 18.14 & 18.46 & 18.46 & 18.78 & 18.78 & 18.94 \\
 & FasterPy & 21.19 & 25.04 & 27.13 & 28.89 & 29.70 & 30.98 & 31.62 & 32.42 \\
 & \textbf{EffiSkill} & 26.48 & 30.34 & 32.91 & 33.55 & 33.87 & 36.28 & 36.92 & 37.40 \\
\midrule
GPT/C++ & Instruction & 31.62 & 37.72 & 40.61 & 42.38 & 44.14 & 45.91 & 47.03 & 48.15 \\
 & RAG & 34.19 & 38.68 & 40.45 & 42.22 & 43.50 & 44.30 & 44.78 & 45.43 \\
 & CoT & 40.93 & 47.19 & 49.60 & 50.40 & 52.33 & 53.45 & 54.09 & 54.74 \\
 & SBLLM & 38.84 & 44.78 & 47.67 & 49.44 & 51.04 & 52.01 & 52.97 & 53.45 \\
 & FasterPy & 38.84 & 43.50 & 45.59 & 46.87 & 48.48 & 49.92 & 50.56 & 51.69 \\
 & \textbf{EffiSkill} & 44.62 & 53.77 & 57.30 & 60.67 & 63.40 & 64.85 & 65.97 & 66.77 \\
\midrule
Qwen/Python & Instruction & 12.52 & 17.01 & 19.58 & 21.03 & 21.99 & 22.95 & 23.92 & 24.40 \\
 & RAG & 7.22 & 11.24 & 13.64 & 16.05 & 17.34 & 18.62 & 19.10 & 19.74 \\
 & CoT & 16.05 & 19.58 & 21.51 & 22.95 & 24.40 & 25.36 & 25.36 & 26.00 \\
 & SBLLM & 6.90 & 8.03 & 8.35 & 8.35 & 8.35 & 8.51 & 8.67 & 8.67 \\
 & FasterPy & 13.96 & 18.30 & 21.83 & 23.92 & 25.84 & 27.61 & 28.25 & 28.25 \\
 & EffiCoder & 14.29 & 15.25 & 15.89 & 16.21 & 16.53 & 16.69 & 16.69 & 16.69 \\
 & \textbf{EffiSkill} & 21.35 & 27.77 & 30.98 & 32.91 & 34.35 & 36.12 & 36.12 & 36.60 \\
\midrule
Qwen/C++ & Instruction & 26.00 & 29.21 & 31.30 & 31.78 & 32.26 & 32.42 & 32.58 & 32.58 \\
 & RAG & 16.37 & 20.55 & 23.11 & 24.56 & 24.88 & 25.04 & 25.04 & 25.04 \\
 & CoT & 23.11 & 25.84 & 27.29 & 27.61 & 28.57 & 29.05 & 29.70 & 30.02 \\
 & SBLLM & 28.41 & 34.19 & 37.72 & 39.65 & 40.45 & 42.05 & 43.02 & 43.98 \\
 & FasterPy & 19.90 & 27.13 & 29.37 & 31.14 & 32.26 & 33.55 & 33.87 & 34.19 \\
 & EffiCoder & 14.77 & 20.39 & 23.11 & 24.56 & 25.20 & 25.20 & 25.20 & 25.20 \\
 & \textbf{EffiSkill} & 34.19 & 44.78 & 48.80 & 52.81 & 54.57 & 55.86 & 56.34 & 56.50 \\
\bottomrule
\end{tabular}
\caption{All 208 OPT@\(k\) values (\%) behind Figure~\ref{fig:topk_growth}. Each setting uses one fixed public-ranked pool.}
\label{tab:all-topk-values}
\end{table*}

\subsection{Speedup distribution and terminal partition}

Figure~\ref{fig:speedup-distribution} assigns every task the best
private-test-correct ratio among the first eight public-ranked candidates.
Missing, incorrect, and non-improving outcomes receive the neutral ratio
1.00.  The five bins are disjoint.
The clipped all-task geometric means for GPT/Python, GPT/C++, Qwen/Python,
and Qwen/C++ are 1.188, 1.448, 1.129, and 1.176.  The corresponding strongest
baseline means are 1.170, 1.432, 1.181, and 1.250.  Thus coverage improves in
all four settings while speedup magnitude does not uniformly improve.

Table~\ref{tab:terminal-counts} partitions all 623 tasks.  ``Private-correct
below threshold'' denotes a selected private-correct candidate with ratio below
\(1.10\times\) and is the largest non-success category.

\begin{table}[H]
\centering
\scriptsize
\setlength{\tabcolsep}{3pt}
\begin{tabular}{@{}lrrrr@{}}
\toprule
Outcome & GPT/Py & GPT/C++ & Qwen/Py & Qwen/C++ \\
\midrule
Success & 233 & 416 & 228 & 352 \\
No public survivor & 8 & 18 & 73 & 69 \\
Public, no private-correct & 4 & 6 & 12 & 11 \\
Private-correct \(<1.10\times\) & 378 & 183 & 310 & 191 \\
Total & 623 & 623 & 623 & 623 \\
\bottomrule
\end{tabular}
\caption{Complete terminal task-count vectors.}
\label{tab:terminal-counts}
\end{table}

\begin{table*}[t]
\centering
\scriptsize
\renewcommand{\arraystretch}{0.84}
\setlength{\tabcolsep}{4.2pt}
\begin{tabular}{@{}llrrrrr@{}}
\toprule
Setting & Method & \(=1.00\) & \((1.00,1.10)\) & \([1.10,1.25)\) & \([1.25,1.50)\) & \(\geq1.50\) \\
\midrule
GPT/Python & Instruction & 10.59 & 57.78 & 17.34 & 4.49 & 9.79 \\
 & RAG & 15.73 & 54.90 & 16.69 & 5.46 & 7.22 \\
 & CoT & 16.53 & 54.74 & 14.45 & 4.49 & 9.79 \\
 & SBLLM & 34.19 & 46.87 & 4.49 & 3.21 & 11.24 \\
 & FasterPy & 18.94 & 48.64 & 18.30 & 4.65 & 9.47 \\
 & \textbf{EffiSkill} & 8.35 & 54.25 & 19.74 & 7.06 & 10.59 \\
\midrule
GPT/C++ & Instruction & 19.90 & 31.94 & 16.37 & 9.63 & 22.15 \\
 & RAG & 22.79 & 31.78 & 15.73 & 6.58 & 23.11 \\
 & CoT & 19.26 & 26.00 & 13.96 & 13.80 & 26.97 \\
 & SBLLM & 21.19 & 25.36 & 15.57 & 9.95 & 27.93 \\
 & FasterPy & 10.75 & 37.56 & 18.78 & 8.51 & 24.40 \\
 & \textbf{EffiSkill} & 9.47 & 23.76 & 23.76 & 11.40 & 31.62 \\
\midrule
Qwen/Python & Instruction & 33.71 & 41.89 & 17.50 & 4.65 & 2.25 \\
 & RAG & 23.43 & 56.82 & 17.34 & 1.93 & 0.48 \\
 & CoT & 27.13 & 46.87 & 18.14 & 4.98 & 2.89 \\
 & SBLLM & 47.03 & 44.30 & 3.85 & 1.61 & 3.21 \\
 & FasterPy & 19.90 & 51.85 & 21.83 & 4.65 & 1.77 \\
 & EffiCoder & 24.08 & 59.23 & 6.42 & 1.77 & 8.51 \\
 & \textbf{EffiSkill} & 19.58 & 43.82 & 27.29 & 5.78 & 3.53 \\
\midrule
Qwen/C++ & Instruction & 40.77 & 26.65 & 7.87 & 5.62 & 19.10 \\
 & RAG & 50.24 & 24.72 & 8.67 & 3.85 & 12.52 \\
 & CoT & 44.62 & 25.36 & 8.51 & 6.10 & 15.41 \\
 & SBLLM & 40.77 & 15.25 & 17.82 & 10.27 & 15.89 \\
 & FasterPy & 50.88 & 14.93 & 11.24 & 8.67 & 14.29 \\
 & EffiCoder & 65.97 & 8.83 & 6.74 & 8.99 & 9.47 \\
 & \textbf{EffiSkill} & 23.11 & 20.39 & 30.50 & 20.55 & 5.46 \\
\bottomrule
\end{tabular}
\caption{All 130 all-task speedup-bin percentages behind Figure~\ref{fig:speedup-distribution}. Rows sum to 100\% up to rounding.}
\label{tab:all-speedup-bins}
\vspace{1ex}

\centering
\begin{tabular}{@{}llrrr@{}}
\toprule
Backbone & Card & Uses & Denominator & Share (\%) \\
\midrule
GPT-5-mini & O001 Constant factors & 2,366 & 10,360 & 22.8 \\
 & O006 Right-size structures & 1,858 & 10,360 & 17.9 \\
 & O018 Cheaper math calls & 1,150 & 10,360 & 11.1 \\
 & O021 Redundant range loops & 1,128 & 10,360 & 10.9 \\
 & O009 Arithmetic to closed form & 572 & 10,360 & 5.5 \\
\midrule
Qwen3-Coder & O001 Constant factors & 1,281 & 6,052 & 21.2 \\
 & O018 Cheaper math calls & 892 & 6,052 & 14.7 \\
 & O006 Right-size structures & 822 & 6,052 & 13.6 \\
 & O029 Rolling-state DP & 622 & 6,052 & 10.3 \\
 & O010 DP state reformulation & 621 & 6,052 & 10.3 \\
\bottomrule
\end{tabular}
\caption{All runtime-linked attachment-frequency values plotted in
Figure~\ref{fig:skill-usage}.}
\label{tab:skill-usage-values}
\end{table*}

Table~\ref{tab:skill-usage-values} contains every value plotted in
Figure~\ref{fig:skill-usage}.  Uses are candidate--card attachment incidences;
the denominator is all such incidences for that backbone.

\endgroup

\section{Optimization Trace: Aizu 1464}
\label{app:examples}

This case study follows one GPT-5-mini/Python task through diagnosis, skill
selection, planning, code generation, public ranking, and private evaluation.
It illustrates how reusable mechanisms are adapted to a program without
replacing an already appropriate high-level algorithm.

\subsection{Diagnosis and skill selection}

The task is Aizu 1464 (\texttt{mixing-solutions}).  The baseline performs a
full sort of the container list inside each of roughly 30 parametric-search
probes, repeatedly normalizes rational values, and carries avoidable
large-integer overhead.  Diagnosis therefore identifies the outer search as
appropriate but its per-probe implementation as the dominant hot path.

Retrieval selects O001 (constant-factor and lightweight-data-structure
optimization) and O021 (redundant range-loop and aggregation refactoring).
These skills jointly target repeated work while preserving the search logic.

\subsection{Selected plan}

The conservative plan retains the parametric-search algorithm and makes four
coordinated changes:

\begin{enumerate}
    \item parse input through buffered tokens;
    \item replace in-place mutation with a local keyed ordering for each probe;
    \item stop the accumulation loop once the required amount is reached; and
    \item keep probe arithmetic integral and perform one final
          \texttt{gcd} normalization instead of constructing repeated
          \texttt{Fraction} objects.
\end{enumerate}

The plan preserves the original input/output interface and exact rational
output format.

\subsection{Mechanism-bearing code changes}

The substantive transformation is shown below; unchanged parametric-search
logic between the displayed hunks is omitted.

\begin{Verbatim}[breaklines=true,breakanywhere=true,fontsize=\scriptsize]
-from fractions import Fraction
+import sys
+import math
 ...
-n, s, c = map(int, input().split())
-alr = [tuple(map(int, input().split())) for _ in range(n)]
+data = sys.stdin.buffer.read().split()
+it = iter(map(int, data))
+n = next(it); s = next(it); c = next(it)
+alr = [(next(it), next(it), next(it)) for _ in range(n)]
 ...
 def value_slope(k):
-    alr.sort(key=lambda x: k*x[2] - (P-k)*x[1])
+    ordered = sorted(alr, key=lambda x: k*x[2] - (P-k)*x[1])
     val = c * s * (P - 2 * k)
     slope = -2 * c * s
     rem = s
-    for a, l, r in alr:
-        z = min(a, rem)
+    for a, l, r in ordered:
+        if rem <= 0:
+            break
+        z = a if a < rem else rem
         val += (k*r - (P-k)*l) * z
         slope += (l+r) * z
         rem -= z
     return val, slope
 ...
-    ans = Fraction(gl*gr*(kl-kr) + gl*fr - gr*fl,
-                   gl-gr)
-ans /= P * M
-print(ans.numerator, ans.denominator)
+    num = gl*gr*(kl-kr) + gl*fr - gr*fl
+    den = gl-gr
+final_num = num
+final_den = den * P * M
+if final_den < 0:
+    final_den = -final_den
+    final_num = -final_num
+g = math.gcd(final_num, final_den)
+print(f"{final_num // g} {final_den // g}")
\end{Verbatim}

The rewrite leaves the asymptotic structure of the parametric search intact.
Its gain comes from reducing allocation, object construction, mutation, and
unnecessary loop work along the repeatedly executed path.

\subsection{Public ranking and private outcome}

The generated candidate passes syntax, interface, and static checks, then
passes every public test and ranks first among public survivors.  It also
passes every held-out private test.  The private aggregate runtimes are

\[
\begin{aligned}
u(S)&=2420.952493740001,\\
u(C1)&=1259.622465,\\
u(S)/u(C1)&=1.921966748775\ldots .
\end{aligned}
\]

Thus the candidate achieves a $1.92\times$ private-test speedup.  The trace
shows the intended division of labor in \tool: diagnosis preserves the sound
outer algorithm, retrieved skills supply reusable local mechanisms, and the
task-specific plan composes them under the program's exact arithmetic and
interface constraints.

\section{Experimental Configuration and Reporting Details}
\label{app:reproducibility}

This appendix consolidates the configuration and reporting details needed to
interpret the study design.  Detailed schemas and algorithms appear in
Appendices~\ref{app:stage1} and~\ref{app:stage2}; all numerical inputs needed
to verify the reported paper claims appear in Appendix~\ref{app:evaluation}.

\subsection{Configuration summary}

\paragraph{Inference compute and APIs.}
LLM inference used hosted APIs; provider-side hardware and FLOPs are
unavailable.  Methods share an at-most-eight-output reporting bound, but calls,
tokens, latency, monetary cost, and total inference compute are not matched.

\subsection{End-to-end evaluation sequence}

For each model--language setting, the evaluation proceeds as follows:

\begin{enumerate}
    \item Freeze the language-specific Operator Skill registry before
          evaluation.
    \item Give every compared method the same canonical input program and task
          statement.
    \item Complete diagnosis, retrieval, planning, generation, and local
          pre-execution processing without candidate-execution feedback.
    \item Generate at most eight candidate outputs.  Missing outputs remain
          unsuccessful positions rather than changing the task denominator.
    \item Execute candidates on the 20 public tests, discard candidates that
          are not public-correct, and sort survivors by aggregate public
          \texttt{runtime\_sum}.
    \item Evaluate the same survivor coordinates on the 80 held-out private
          tests.  Private outcomes never alter the public ordering.
    \item Compute OPT@$k$ over all 623 tasks for each prefix
          $k=1,\ldots,8$.
\end{enumerate}

This ordering prevents private-test feedback from entering candidate creation
or selection while allowing public tests to provide the same correctness and
ranking role for every method.

\begin{table}[H]
\centering
\scriptsize
\begin{tabular}{@{}p{.27\columnwidth}p{.67\columnwidth}@{}}
\toprule
Component & Configuration \\
\midrule
Mining corpus & 900 unique slow/fast pairs per language after canonicalizing
1,800 directional rows; C++ from PIE; Python from PIE, Mercury, DeepMind Code
Contests, and CodeParrot APPS; separate from EffiBench-X evaluation tasks \\
Stage-I LLM actions & GPT-5.1 for structured trace extraction and cluster
summarization \\
Signature & transformation type, complexity change, triggers, steps,
bottlenecks, and problem tags \\
Stage-I clustering & normalized TF--IDF plus
\texttt{all-MiniLM-L6-v2}, cosine silhouette selection, and KMeans seed 42 \\
Skill library & separate within-language registries; 29 Python Operator Skills
in the library analysis \\
Stage-II models & \texttt{gpt-5-mini-2025-08-07} or
\texttt{Qwen3-Coder-30B-A3B-Instruct}, depending on the experimental setup \\
Candidate bound & at most eight generated outputs per task and method \\
Selection boundary & generation and static processing first; public tests
filter and rank; private tests report correctness and speedup \\
Task split & 623 tasks per language; 20 public and 80 private tests per task \\
Success criterion & private correctness and at least $1.10\times$ speedup
within the first $k$ public-ranked survivors \\
Statistical analysis & 10,000 one-sided paired task-bootstrap replicates;
percentile 95\% intervals; Holm adjustment across eight endpoints \\
\bottomrule
\end{tabular}
\caption{Study-configuration summary.}
\label{tab:repro-checklist}
\end{table}

\subsection{Metric reconstruction}

For task $q$, let $S_q$ be the canonical program and
$c_{q,1},\ldots,c_{q,m_q}$ the public-correct candidates in increasing public
runtime order.  The task succeeds at prefix $k$ exactly when

\[
\begin{aligned}
\exists j&\leq\min(k,m_q):\ \mathrm{pass}_{\rm priv}(c_{q,j}) \\
&{}\land u_q(S_q)/u_q(c_{q,j})\geq1.10.
\end{aligned}
\]

OPT@$k$ is 100 times the mean of this binary indicator over all 623 tasks.
Consequently, a task with no public survivor contributes zero, later values of
$k$ expose additional candidates from the same fixed ordering, and all
methods retain the same task denominator.

The full endpoint margins, confidence intervals, adjusted $p$-values, and
explicit bootstrap seeds are reported in Table~\ref{tab:full-statistics}.
Table~\ref{tab:ablation} reports the ablation values, and
Table~\ref{tab:terminal-counts} provides the integer task counts underlying
the terminal-outcome percentages.

\FloatBarrier

\end{document}